Northeastern University
Boston, Mass 02115

# Multiplicity Distributions in $b \to s\ gluon$ Decays

**John Swain**

**Department of Physics**
**Northeastern University**
Boston, MA 02115, USA

**Abstract**

The final states for the process $b \to s\ \gamma$ have been extensively discussed in the literature. Similarly-detailed analyses for the case $b \to s\ gluon$ have not been performed. Generally this process is searched for in 2-body decays such as $B^0 \to K^+\pi^-$. We present simple arguments to suggest that most of the time the quark-level process will give rise to final states with rather high multiplicities. Comments are made about the applicability of these results to $b \to d\ gluon$ and hadronic $b \to u$ decays.

Since the process $b \to s\gamma$ was first considered by Campbell and O'Donnell in 1982 [1], there have been many discussions of how this quark level process is expected to be manifest in various final states. The most complete analysis to date is that of Altomari [2], though there are a number of papers concentrating on the probability of the $K^*\gamma$ final state. This topic has become especially interesting since the CLEO observation of $b \to s\gamma$ in the decay $B \to K^*(892)\gamma$ [3]. This measurement is consistent with a branching ratio for the quark level process $b \to s\gamma \sim 10^{-4}$ with 5%-40% probability of this appearing in the $K^*(892)\gamma$ final state.

The related process $b \to s$ *gluon* is also expected, but with a much larger branching ratio, perhaps as high as 1% [5] (for a general review of $b \to sX$ amplitudes see reference 6 and references therein). Supersymmetry [7] or an extended Higgs sector [8] can raise this fraction dramatically. Given the large number of $B$ meson decays available for study in the ARGUS, CLEO, and LEP data samples, one might be encouraged to look then for this process as well. The usual strategy proposed has been to search in final states such as $B \to X_s X$ where $X_s$ is a strange meson, and $X$ is $\pi$, $\phi$, $\rho$, etc. Each exclusive final state is expected with only tiny branching ratio of $\sim 10^{-5} - 10^{-4}$ [9]. This then prompts the obvious question of which final states are expected to have large branching ratios. In this paper it is argued that the process $b \to s$ *gluon* can be reasonably modelled as one in which a B meson is replaced by a collection of light quarks and gluons with the same total energy, which then fragments into a large number of final state particles. The physical picture is essentially that the weak decay replaces the b quark by an s quark and dumps on the order of 5 GeV of energy into the QCD vacuum.

Consider a $B$ meson made of a $b$ quark and a light antiquark $\bar{q}$ $(q = u$, or $d)$ bound together by the strong force. Now consider the quark level process $b \to s$ *gluon*. This can happen, following Grigjanis et al. [10], in essentially four different ways with representative diagrams illustrated in figure :

1. $b \to s\ gluon$ :    *gluon* is radiated, $\bar{q}$ is a spectator

2. $b\ \bar{q} \to s\ \bar{q}$ :    gluon is spacelike and cannot be emitted freely

3. $b \to s\ gluon\ gluon$ :    *gluon* couples to a pair of gluons through the QCD 3-gluon vertex, or gluons couple directly to a quark line (process not shown in diagram), $\bar{q}$ is a spectator

4. $b \to s\ q'\ \bar{q}'$ :    *gluon* couples to $q'\bar{q}'$ through the QCD gluon-quark-quark vertex, $q'$ is a light quark, $\bar{q}$ is a spectator

As shown in [10], processes 3 and 4 are expected to dominate if the decays occur within the framework of the Standard Model. This is not necessarily true in extended models [7] where other particles such as squarks might appear in the loops.

In the absence of techniques which can handle the fragmentation and hadronization from first principles, we model these using the JETSET 7.3 Monte Carlo [11,12] string fragmentation model.

In more detail, the four processes described above are simulated as:

1. a 3-particle jet composed of a $\bar{u}$ quark at rest in the rest frame of the initial B meson, and a gluon and a strange quark produced with momenta appropriate for products of a 2-body decay of a b-quark of constituent mass $m(b) = 5.0\mathrm{GeV/c^2}$. The mass of the strange quark is taken to be $m(s) = 0.5\ \mathrm{GeV/c^2}$.

2. the result of an $e^+e^-$ collision producing an $s\bar{u}$ pair with an energy $m(b) - m(s)$. This process is, of course, forbidden in the Standard Model at tree level, and in fact cannot be directly generated in the JETSET 7.3 model. It can, however, be simulated by using the process $e^+e^- \to q\bar{q}$ and then replacing the $q$ quark by an $s$ quark, and readjusting the energy appropriately before allowing fragmentation to occur. The object which is allowed to decay essentially looks like a highly-excited $K^*$ meson.

3. a 4-jet event where the $b$ quark is allowed to decay into an $s$ quark and 2 gluons with momentum distributions from 3-body phase space and the $\bar{u}$ quark is at rest in the rest frame of the initial B meson.

4. the preceding case, but with the gluons replaced by $u\bar{u}$ or $d\bar{d}$ pairs with equal probability.

These four models of the $b \to s\ gluon$ process are allowed to fragment and the particles in the final states are examined. Results of $2 \times 10^5$ Monte Carlo events for each model, with a 50/50 mixture of $B$ and $\bar{B}$ mesons, half of which were charged and half of which were neutral, gave the charged multiplicities shown in Figure . $K_s^0$ and $K_L^0$ are considered as neutral hadrons, and not decayed to pions. The total hadronic multiplicities are shown in Figure .

We see from this that the results of the simulations of the 4 basic processes are very similar. We also see that the final states expected from $b \to s\ gluon$ decays should be, for the most part, states of high multiplicity, containing one kaon and more than 3 pions. The fraction of decays into K$\pi$ and K$\pi\pi$, which are the channels experimentally searched for to date, are together expected to make up less than 10% of the total branching ratio for the quark level process. The distributions are given as histograms as they do not fit well with simple distributions such as Gaussian or negative binomial.

The JETSET 7.3 model, as used here, is quite stable against perturbations of parameters, and there is no fine-tuning done in this study. In principle one could try to extract estimates for the relative importance of various resonances, but these would be more difficult to justify, since they are incorporated into the JETSET model via phenomenological parameters which must be put in by hand. We note here that the string fragmentation model has difficulties in describing baryon production, and while we expect that baryonic final states will not constitute a major fraction of the decay modes, they certainly provide an interesting area for further study. A more complete treatment would also include the effects of the appropriate matrix element for the quark-level $b \to s$ process, and account for the Fermi motion of the quarks in the B meson.

Such extensions should mainly have the effect of distorting the relative fractions of energy carried by the light quarks and gluons before fragmentation. The point of this paper, however, is not to make precise calculations of the decay rates into various final states, but simply to point out that on very general grounds, almost independent of the details of the processes that take place, one would expect high multiplicity final states. It is important to note that the kinematics of the 4 basic processes generated according the phase space are quite different, and yet the multiplicity distributions are quite similar, so we expect that the simplifications in the model as described above are unlikely to modify the conclusions in any substantial way.

A cross-check can be made by examining data on charged-particle multiplicity in $e^+e^-$ interactions. At energies near 5 GeV (corresponding to the mass of a $B$ meson), the data in [13] are consistent with the average charged multiplicities for the processes under consideration in this paper.

From the experimental point of view, it should be noted that high multiplicity decays are expected not only directly from $b \to s\ gluon$, but also as end-products of charm decays [14]. Decays chains such as $B \to D^{(*)} + n\ pions$ followed by $D^{(*)} \to K + n'\ pions$ will give rise to the same sorts of final states. The separation of decays due to these two processes is expected to be rather difficult. If the branching ratio for $b \to s\ gluon$ is sufficiently large, it might be reflected in an enhancement of such high multiplicity final states. More likely, it will be necessary to use cuts against charm contamination such as rejection of subsystems of the $K + n - pion$ system with invariant masses close to those of known charm mesons. If the experimental setup were such that beauty and/or charm decay vertices could be resolved, there would be a much-improved chance to reduce background and obtain better limits, or perhaps observe a signal.

We also note in passing that similar conclusions apply to the related process where the $s$ quark is replaced by a $d$ quark, though this is expected to be greatly suppressed by CKM matrix elements relative to $b \to s\ gluon$ and hence is of less interest experimentally. One might expect to find similar conclusions for hadronic $b \to u$ decays, but this is not completely straightforward since one might expect the off-shell $W$ to couple often to a $\rho$ or $\pi$ meson which could then carry off enough energy to reduce the multiplicity for the remaining system. Still, one would expect on general grounds again that a large fraction of the total rate will be into states with high multiplicities, though not as high as those for $b \to s\ gluon$.

I would like to thank Bruce Campbell for periodically, over the last 3 years or so, urging me to write this paper, and Lucas Taylor for helpful discussions and a critical reading of a draft of this Letter.

## Figure Captions

**Figure 1:** Processes 1,2,3,4 (clockwise from the upper left) for $b \to sg$.

**Figure 2:** Probabilities of charged multiplicities for processes 1,2,3, and 4.

**Figure 3:** Probabilities of hadronic multiplicities for processes 1,2,3, and 4.

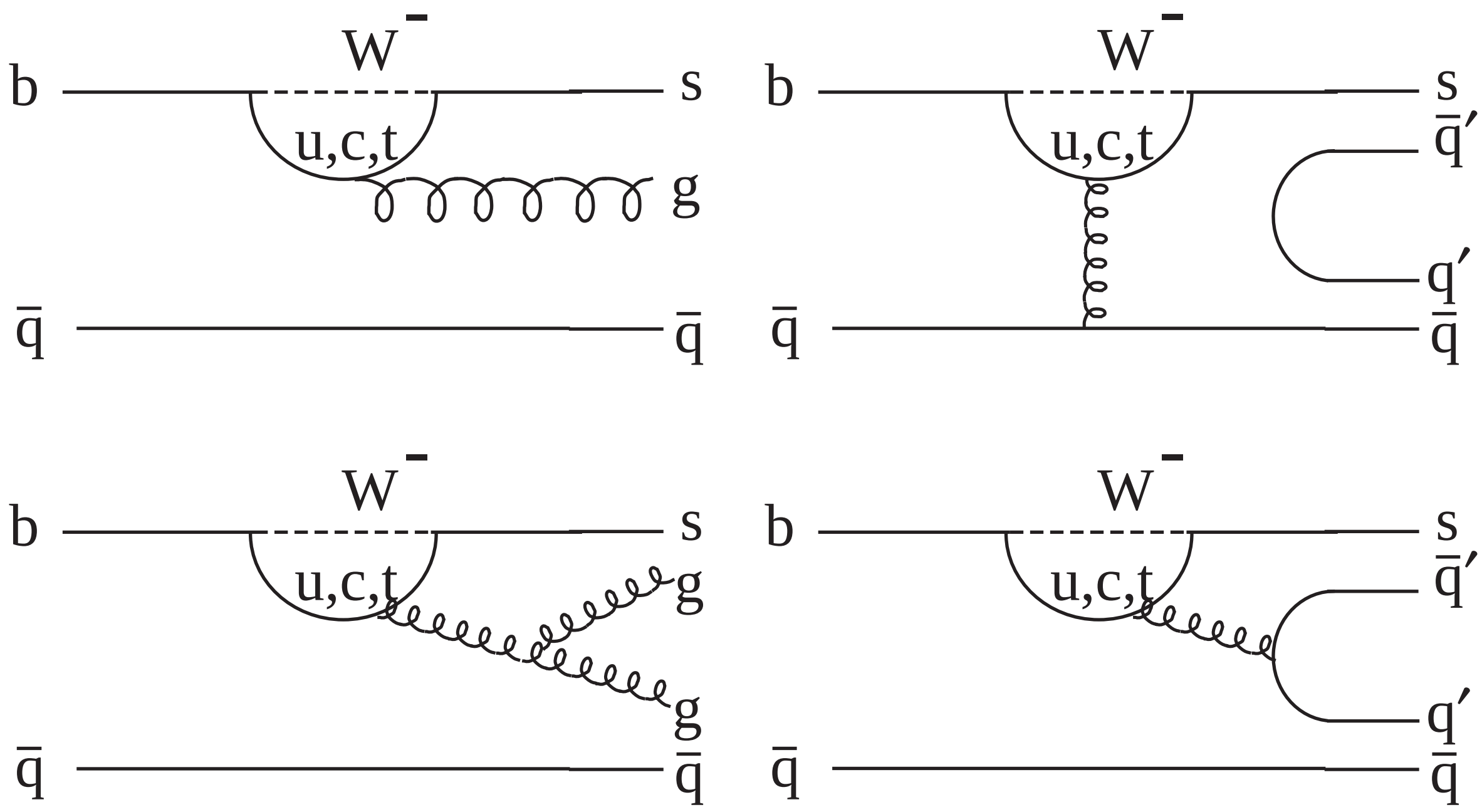

**Figure 1.**

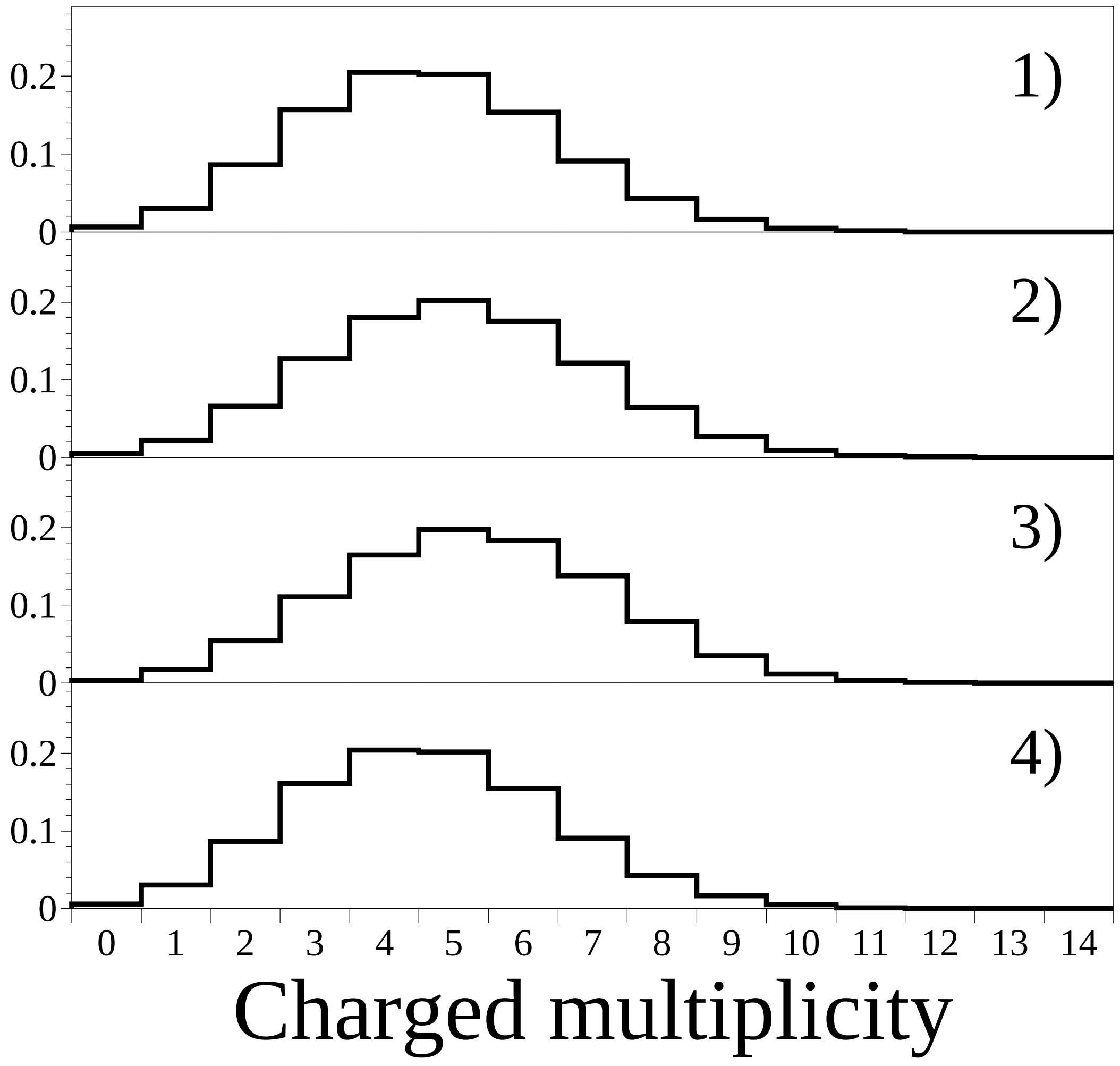

**Figure 2.**

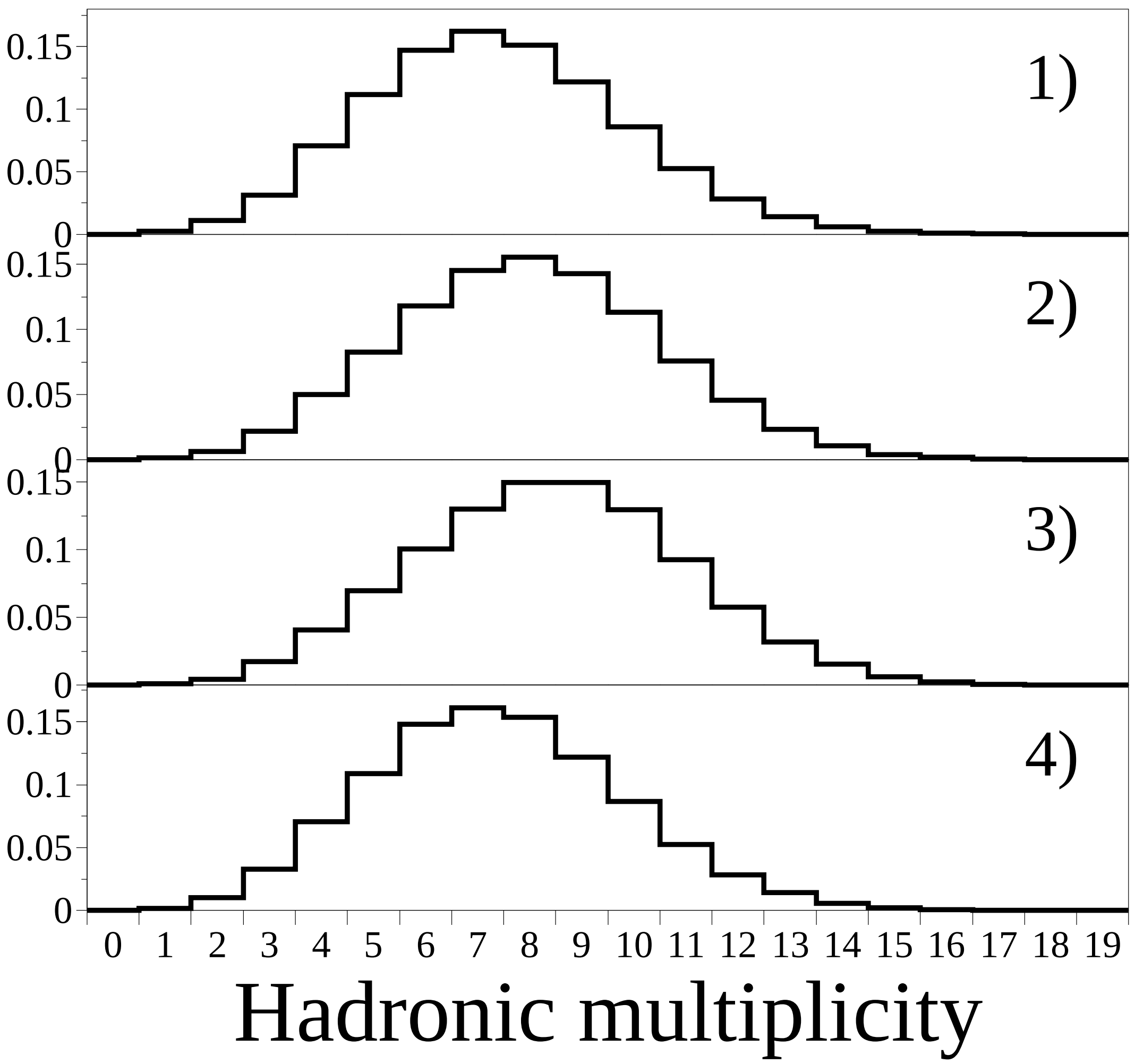
0.15
0.1
0.05
0
1)
2)
3)
4)
0 1 2 3 4 5 6 7 8 9 10 11 12 13 14 15 16 17 18 19
Hadronic multiplicity

**Figure 3.**